\documentclass[twocolumn,showpacs,preprintnumbers,amsmath,amssymb]{revtex4}

\usepackage{amsmath}
\usepackage{graphicx}
\usepackage{amssymb}
\usepackage{amsfonts}
\usepackage{latexsym,epsfig}

\def\beq{\begin{eqnarray}}
\def\eeq{\end{eqnarray}}

\def\al{\alpha}

\def\ga{\gamma}

\def\na{\nabla}

\begin{document}


\title{Torsion Discovery Potential and Its Discrimination at CERN LHC}
\vskip 6mm
\author{F.~M.~L.~de Almeida Jr., A.~A.~Nepomuceno}
\affiliation{Instituto de F\'isica, \\
Universidade Federal do Rio de Janeiro,\\
Rio de Janeiro, RJ, Brazil}
\email{ asevedo@cern.ch,marroqui@if.ufrj.br}
\author{M.~A.~B. do Vale}
\affiliation{Departamento de Ci\^encias Naturais,\\ 
Universidade Federal de S\~ao Jo\~ao del Rei, \\
S\~ao Jo\~ao del Rei, 36301-160, MG, Brazil }
\email{aline@ufsj.edu.br}

\vskip 12mm

\begin{abstract} 
Torsion models constitute a well known class of extended quantum gravity models.
In this paper we study some phenomenological consequences of a torsion field interacting with  fermions at LHC. A torsion field could appear as a new heavy state characterized by its mass and  couplings to fermions.  These new states will form a resonance decaying into difermions, as occurs in many extensions of the Standard Model, such as models predicting the existence of additional neutral gauge bosons, usually named $Z^\prime$.  Using the dielectron channel we evaluate the integrated luminosity needed for a $5\sigma$ discovery as a function of the torsion mass, for different coupling values.
We also calculate the luminosity needed for discriminate, with 95\% C.L., the two possible different torsion natures. Finally, we show that the observed signal coming from the torsion field could be distinguished from a signal coming from a new neutral gauge boson, provided there is enough luminosity.
\end{abstract}
\pacs{13.85.-t,04.60.-m}

\maketitle

\section{Introduction}
One great challenge of the Standard Model is the difficulty to incorporate gravity. Several  alternative theories of quantum gravity, extending general relativity, introduce an extra field called torsion~\cite{Hehl,Shapiro,Hammond}, where the spin of the elementary particles is the source of the torsion. 
In a phenomenological point of view, torsion can be treated as a fundamental propagating field, either of vectorial or pseudo-vectorial nature,
characterized by its mass and the coupling values ($\eta_f$) between torsion and fermions.
Despite there are strong theoretical restrictions on the possible parameters of the propagating torsion field~\cite{Guilherme}, there is still room for detecting this field on high energy colliders.
The introduction of torsion fields leads to  singularity-free cosmological models~\cite{Gasperini,Puetzfeld} and they can be
investigated in the near future high energy particle colliders, such as the LHC at CERN, through their interaction with the fermions. Some torsion studies have been previously performed for LEP and TEVATRON~\cite{Belyaev,Mahanta} and even for the LHC~\cite{Aline}.  Other torsion aspects have been also recently studied~\cite{Kostelecky,Kruglov}.
The LHC, which will have a center-of-mass energy of 14 TeV, will provide  unprecedented conditions to investigate many alternative models and will extend the search for new heavy particles up to the 5-6 TeV range, depending on the integrated luminosity obtained. The machine must start  its activity at low luminosity ($\cal L = $ $10^{-33}cm^{-2}s^{-1}$), reaching high  luminosity ($\cal{L} = $ $10^{-34}cm^{-2}s^{-1}$) after one year of operation~\cite{Smith}. 

Here we present the discovery potential of torsion field at the LHC. The dielectron channel is used 
to calculate the luminosity needed for a 5$\sigma$ discovery as a function of the torsion mass and the fermion-torsion vertex coupling.  We consider all the couplings $\eta_f$ equal.
The analysis with early LHC data of new heavy states formed by opposite sign dileptons  will have great importance, since these
signatures are included in many Standard Model extensions. The LHC experimental data can eliminate or corroborate some of these proposed phenomenological extensions.

In a second part of this work, we present the luminosity needed for the discrimination of the torsion field nature, vectorial or pseudo-vectorial, with 95\% C.L.. Finally, we present a variable capable of separating the torsion field from a new vector neutral gauge boson ($Z^\prime$), predicted by several models, such as the $E_6$~\cite{Rizzo}. We show that, once a signal of a dilepton resonance is experimentally found, the accumulated asymmetry (defined in Sec. V) can discriminate between the torsion and the $Z^\prime$. The parameters and particle efficiencies of the ATLAS detector have been roughly used in  this paper, although no detector simulation had been performed.

\section{The Torsion Field}

There are many experimental and theoretical arguments saying that the minimal Standard Model (SM) is not a complete theory and it should be extended.
One possibility is to extend it in such way to include also gravity.
This inclusion should predict the observable value of the cosmological constant and bring new observable effects in high energy elementary particle physics.
The SM is described using three types of fields: scalar, vectorial and spinorial, by the other hand general relativity has in addition the metric field which describes the geometric property of space-time.
A strong candidate to be included in the SM is the space-time torsion field which adds some independent characteristics to the space-time geometry as is shown in the general relativity with torsion ~\cite{Shapiro}.
The torsion field can be considered as being composed of three irreducible components: 
the axial vector $S^\mu$, the vector trace $T_\alpha$, and the 
tensor $q^{\alpha}_{. \beta \mu}$.

The general minimal and non minimal action of a Dirac fermion coupled to torsion can be written for practical purposes as ~\cite{Aline}

\beq
S_f \,=\, \int d^4x \sqrt{g}\,\{\,i\bar{\psi}\gamma^\mu 
\big( \na_\mu - i \eta_1\gamma^5S_\mu  \nonumber \\
+ i\eta_2T_\mu\big)\psi 
 - m\bar{\psi}\psi \}
\label{t2}
\eeq

\noindent
where $\eta_1$ and $\eta_2$ are two dimensionless parameters and $\na_\mu$ is the
Riemannian covariant derivative (without torsion).  The $q^\alpha_{.\beta \mu}$ tensor decouples completely from the fermion fields for the minimal and non minimal actions.

One has two approaches: the minimal and the non minimal interaction.  The minimal interaction gives $\eta_1= -1/8$ and the vectorial torsion part decouples completely from the spinor fields. This case is equivalent to put $\eta_2 =0$ in Eq. 1 and then one has a pure axial vector coupling.  So for the minimal case the theory gives a fixed value of $\eta_1$ which corresponds to a pure axial vector coupling ~\cite{Shapiro}.

By the other hand, the non minimal interaction allows both $\eta_1$ and $\eta_2$ to have non zero values.  In order to simplify our study and analyze the extreme cases, we are considering the pure axial vector ($\eta_f = \eta_1 \neq 0$ and $\eta_2 =$0) and the pure vectorial ($\eta_1 = $0 and $\eta_f = \eta_2 \neq 0$) cases, although the mixed case ($\eta_1 \neq 0$ and $\eta_2 \neq 0$) is not at all excluded.  We analyze the non minimal case since the minimal one is very restrictive due to the fixed values of $\eta_1$ and $\eta_2$ in Eq. 1.

For the axial case, the non minimal interaction of torsion with Standard Model fermions will be given by the following action, where $\psi_{(i)}$ stands for each of the fermions:

\beq
{\cal S}_{A-non-min}^{TS-matter} 
\,=\, i\,\int d^4x\sqrt{g}\,\,{\bar \psi_{(i)}}\,
\Big(\, \ga^\al\,\na_\al \nonumber \\
+ i\eta_{1i}\ga^5 \ga^\mu S_\mu
- im_i \,\Big)\,\psi_{(i)}
\label{dirac2-nm1}
\eeq

While for the vectorial case, the non minimal action will be:

\beq
{\cal S}_{V-non-min}^{TS-matter} 
\,=\, i\,\int d^4x\sqrt{g}\,\,{\bar \psi_{(i)}}\,
\Big(\, \ga^\al\,\na_\al \nonumber \\
- i\eta_{2i} \ga^\mu T_\mu
- im_i \,\Big)\,\psi_{(i)}
\label{dirac2-nm2}
\eeq

\noindent
where $\eta_{1i}$ and $\eta_{2i}$ are the interaction parameters for the corresponding fermion spinor and the subscripts V and A stand for the vectorial and pseudo-vectorial cases.
From the theoretical point of view  the nonminimal interaction with the geometric vector component of torsion $T_\mu$ can not be ruled out ~\cite{Shapiro} and therefore it is worthwhile to work out it in the same manner as it is done with the axial vector component. 

We have considered the non minimal parameters equal for all fermions, contrary of what was done on \cite{Aline} where the top quark coupling was taken to be different from the others due to its larger mass because of the renormalization group running. So, from now on, $\eta_{f} = \eta_{1i} = \eta_1$, if $\eta_{2} = 0$ and $\eta_{f} = \eta_{2i} = \eta_2$, if $\eta_{1} = 0$ then we are dealing with two parameters only, $\eta_f$ and $M_{TS}$ for the axial and vectorial cases.
We are studying the channel $pp \rightarrow e^+ e^-$ and this assumption with respect to the top quark coupling makes practically no difference in the results here obtained.  This assumption changes slightly the torsion width.

\section{Event Selection}

Torsion (TS) described as pure vectorial and pure pseudo-vectorial field was implemented in CompHep event generator~\cite{Comphep}. Using this package, we simulate the process $pp \rightarrow \gamma/Z/TS \rightarrow e^{+}e^{-}$ with a center-of-mass energy of $14$ TeV and with full interference between the torsion and the SM gauge bosons. The events were generated for three different values of the torsion mass: $1$, $3$, and $5$ TeV and for three values of $\eta_f$: $0.2$, $0.5$ and $0.8$, which obeys the limits obtained by~\cite{Aline} and the constraint $M_{TS}/\eta_f \gg m_f$ (or, equivalently, $M_{TS} \gg \eta_{f} \times m_{f}$), obtained in~\cite{Guilherme} for the pure axial case.  For the pure $T_\mu$ case, there are no such theoretical limits. A cut on the dielectron invariant mass of $500$ GeV was applied on generated events in order to reduce the SM background. The CTEQ6L parton distribution functions were employed. The pseudo-rapidities of the electron and the positron were required to be $|\eta| < 2.5$ and each final particle energy should be greater than $5$ GeV, both conditions due to the ATLAS detector characteristics. Since in a more realistic scenario jets can be misidentified as electrons in calorimeters, which affect the electron identification, we assume a dielectron identification efficiency of 62\%. The overall efficiency, which take into account the cut on $\eta$ and the electron identification is around 50\%~\cite{Atlas}. New neutral gauge models based on $E_6$ were also implemented in CompHep.

\section{Torsion Discovery Potential at LHC}

\begin{figure} 
\begin{center}
\includegraphics[width=.5\textwidth]{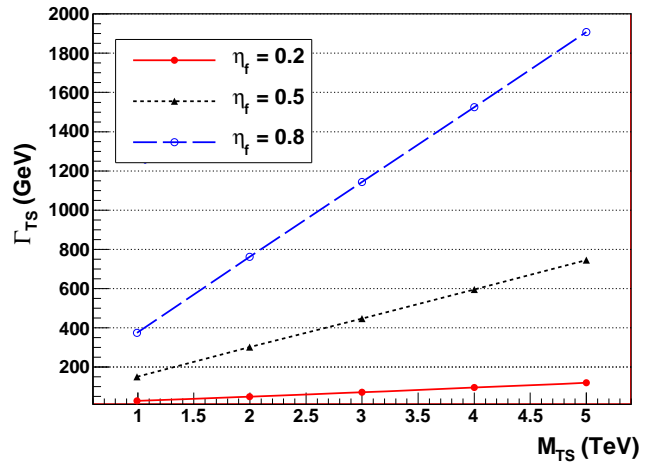}
\caption{Torsion widths as a function of torsion mass for several fermion-torsion couplings ($\eta_f = 0.2, 0.5$ and 0.8).}
\label{fig:figure01}
\end{center} 
\end{figure}

\begin{figure} 
\begin{center}
\includegraphics[width=.5\textwidth]{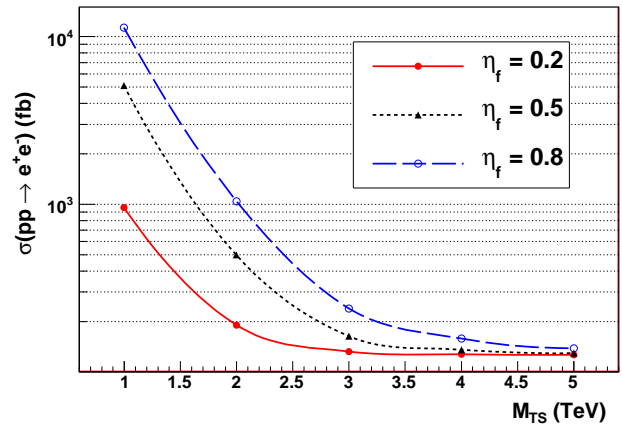}
\caption{Torsion production cross section as a function of torsion mass, for several 
fermion-torsion couplings ($\eta_f = 0.2, 0.5$ and 0.8).}
\label{fig:cross_section}
\end{center} 
\end{figure}

In Fig.~\ref{fig:figure01} we present the torsion width as a function of its mass for several fermion-torsion couplings.  Due to its huge width, in special for large $\eta_f$ values, torsion interaction can not be approximated as a contact interaction.  Figure~\ref{fig:cross_section} shows the production cross section at NNLO for the process $pp \rightarrow \gamma/Z/TS \rightarrow e^{+}e^{-}$ as a function of torsion mass and for three different values of $\eta_f$. LO cross section obtained with the CompHep package was multiplied by a constant factor $K = 1.35$ in order to take into account the next-to-next-leading order QCD corrections. As expected, the cross section grows as $\eta_f$ increases, since the fermion-torsion coupling becomes stronger.  We can see that for $M_{TS}$ above 4 TeV, there is no significant difference between the cross sections for the different couplings, showing clearly the effect of the parton distribution functions. The widths and total cross-sections are equal for both the pure axial and the pure vectorial cases.

The dielectron channel was chosen to calculate the torsion discovery potential at LHC since ATLAS has a better electron resolution when compared to muon~\cite{Atlas2}.  A signal of a torsion field interacting with the fermions should be observed above the neutral Drell-Yan (DY) process that is the main background in the search of new heavy dilepton resonances. The statistical significance of the torsion signal in the presence of background can be obtained using the likelihood-ratio estimator 

\begin{equation}
S_{L} = \sqrt{2 \textnormal{ln} (L_{s+b}/L_{b})}
\end{equation}

\noindent
where $L_{s+b}$ represents the maximum likelihood value obtained from the unbinned likelihood fit to the dielectron invariant mass assuming that a signal is presented in the data, while $L_{b}$ is the maximum likelihood value obtained with a background only hypothesis. It has been shown that this estimator has the desired feature of not having an overestimated or underestimated probability of a false discovery~\cite{cousin05}. 
 
The probability density function (pdf) used to fit the observed dielectron invariant mass spectra is given by~\cite{cousin06}. 

\begin{equation}
p\,(M_{ee}) = f_{s} p_{s}\,(M_{ee}) + (1 - f_{s}) p_{b}\,(M_{ee})
\end{equation}

\noindent
where $p_s$ is the pdf of the signal, given by a Breit-Wigner distribution, and  $p_b$ is the pdf of the background, modeled as exponential functions. The parameters to be fitted are the signal fractions $f_s = N_s/(N_s + N_b)$ (where $N_{s (b)}$ are the number of signal (background) events), the torsion mass $M_{TS}$ peak position and the torsion width $\Gamma_{TS}$. 

The pdfs that describe the background shape depends on the invariant mass region, and are given by 

\begin{eqnarray}
p_{b}\,(M_{ee}) \propto \left\{\begin{array}{ll}
exp(-k_1 M^{0.2}_{ee}), \: \textnormal{if} \; 500 < M_{ee} < 3 \, \textnormal{TeV} \\
\\
exp(-k_2 M_{ee}), \: \textnormal{if} \: M_{ee} > 3 \, \textnormal{TeV}

      \end{array}
\right.
\end{eqnarray}

\begin{figure} 
\begin{center}
\includegraphics[width=.5\textwidth]{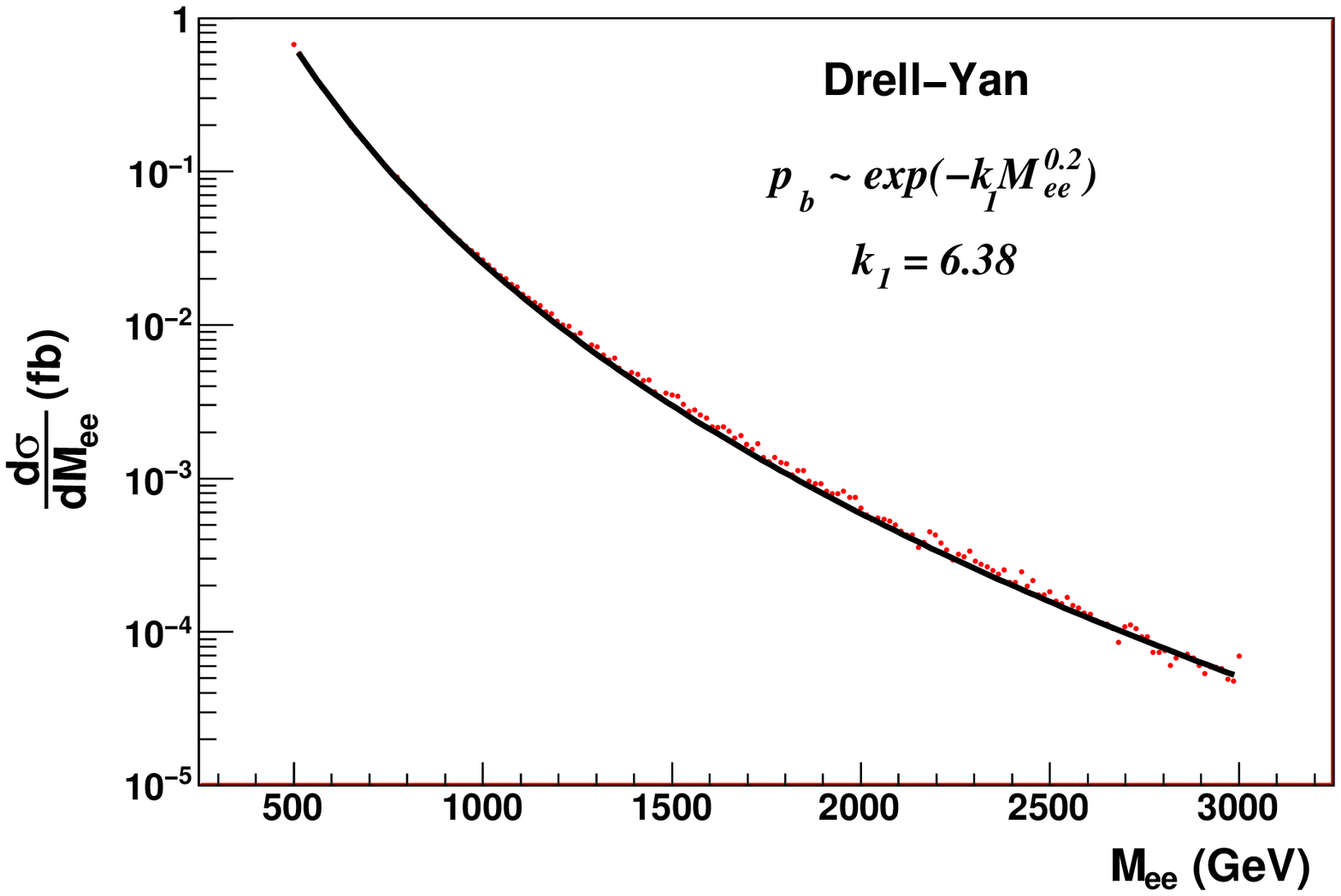}
\caption{Background shape determination for $M_{ee} < 3$ TeV}
\label{fig:back1}
\end{center} 
\end{figure}

\begin{figure} 
\begin{center}
\includegraphics[width=.5\textwidth]{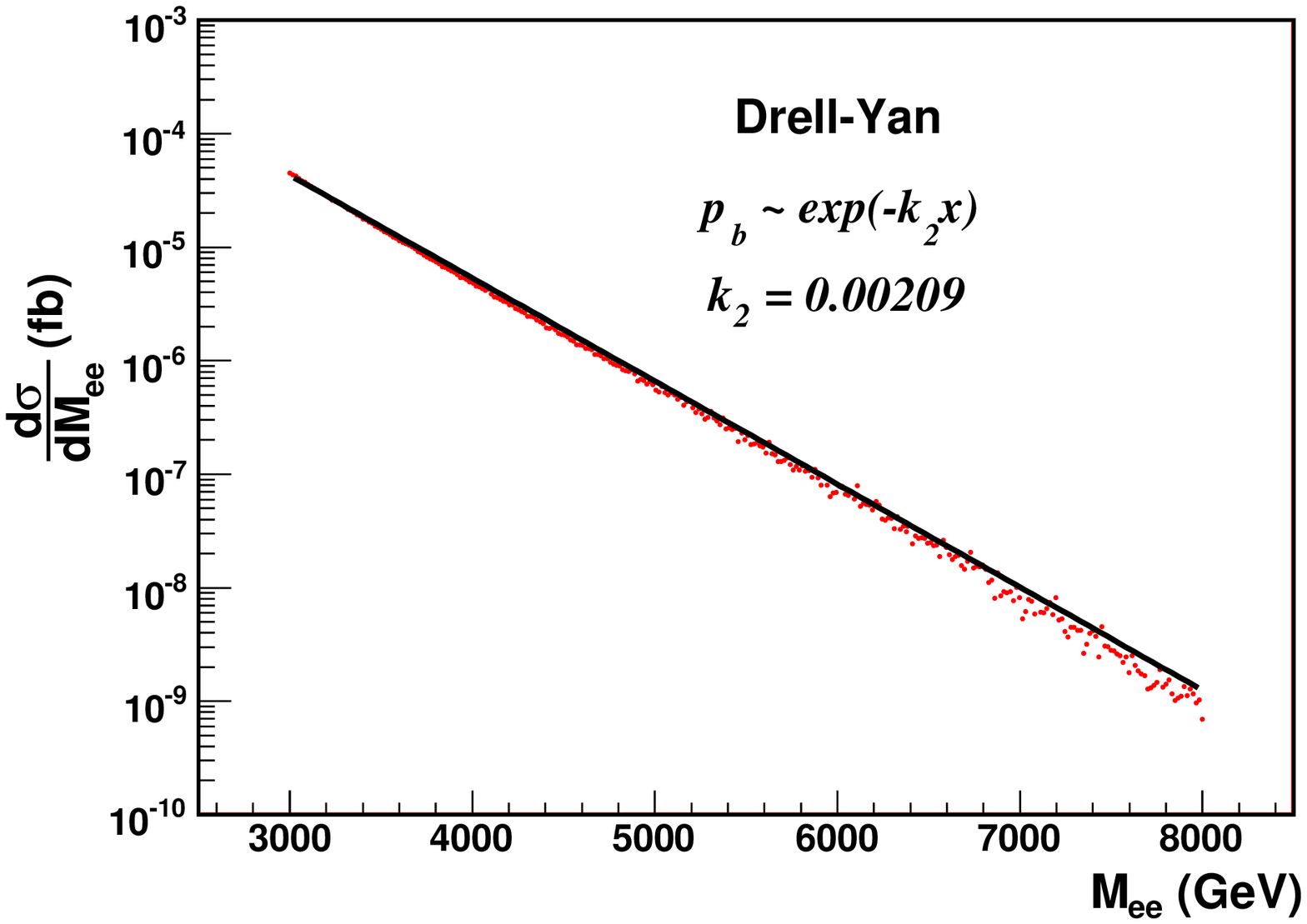}
\caption{Background shape determination for $M_{ee} > 3$ TeV}
\label{fig:back2}
\end{center} 
\end{figure}

The values of the parameters $k_1$ and $k_2$ are obtained from fits to the DY invariant mass Monte Carlo (MC) distribution in the full mass region of interest. Figures ~\ref{fig:back1} and ~\ref{fig:back2} show the fits and the resultant values of $k_1$ and $k_2$: $k_1 = 6.38$ and $k_2 = 0.00209$. We can see that the DY background can be well described by the proposed parametrization.

\begin{figure} 
\begin{center}
\includegraphics[width=.5\textwidth]{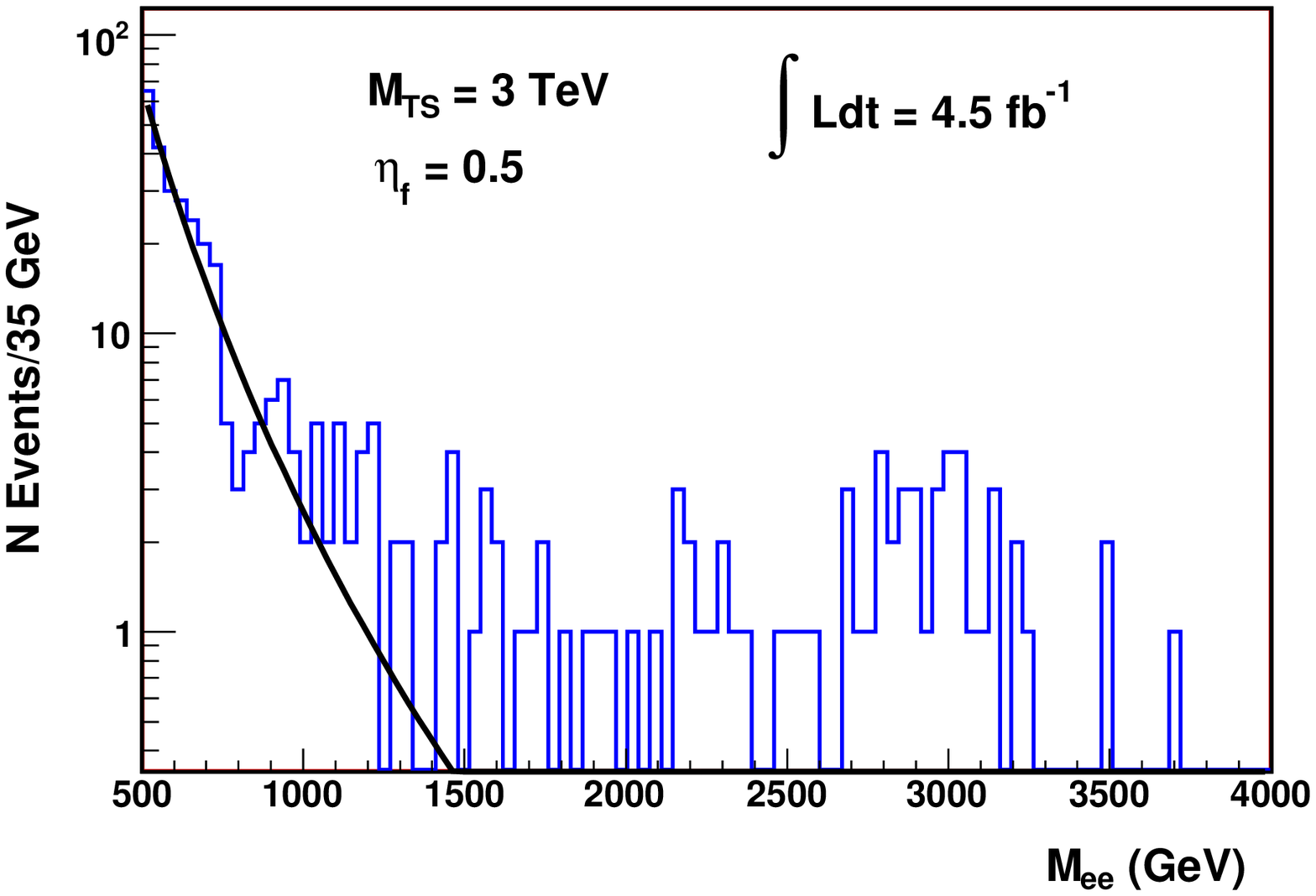}
\caption{Mass cut determination for $M_{TS} = 3$ TeV}
\label{fig:cut3tev}
\end{center} 
\end{figure}

\begin{figure} 
\begin{center}
\includegraphics[width=.5\textwidth]{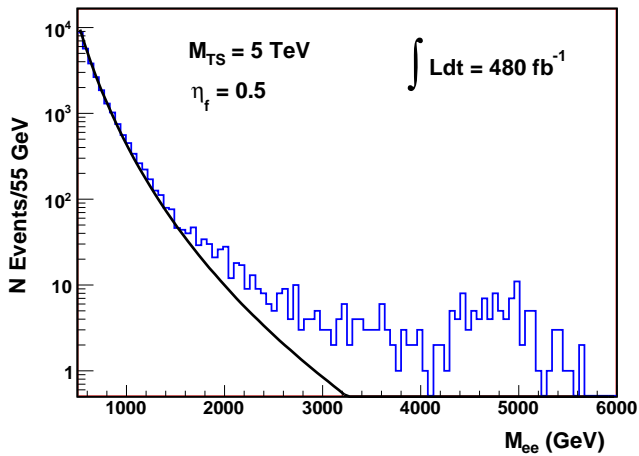}
\caption{Mass cut determination for $M_{TS} = 5$ TeV}
\label{fig:cut5tev}
\end{center} 
\end{figure}

In order to increase the signal-to-background ratio, it was applied a $M_{ee}$ cut to eliminate as much as possible the background contamination on the mass signal. Figures ~\ref{fig:cut3tev} and ~\ref{fig:cut5tev} illustrate for two different values of torsion masses how the cut is obtained: the Equation 6 is used to fit the invariant mass distribution, with $k_{1,2}$ fixed at the values mentioned above, and with the amplitude been the only fitted parameter. The $M_{ee}$ cut is defined as the point where the $d\sigma/dM_{ee}$ reaches the value $0.5 fb$. Table ~\ref{tab:tab1} shows the cut values obtained for the different torsion masses and couplings considered in this study. 

\begin{table}
  \begin{center}
    \caption[]{%
      Cuts on $M_{ee}$ obtained from the exponential fit to the invariant mass for different values of $M_{TS}$ and $\eta_f$.
    \label{tab:tab1}}
\begin{tabular}{|c|c|c|c|}
\hline
&\multicolumn{3}{|c|}{Torsion Mass $(M_{TS})$} \\
\hline
$\eta_f$ & $1 \, \textnormal{TeV}$ & $2 \, \textnormal{TeV}$ & $5 \, \textnormal{TeV}$ \\ \hline
     $0.2$    &     $660$   &   $1570$  &   $3500$  \\ \hline
     $0.5$    &     $680$   &   $1360$  &   $3300$  \\ \hline
     $0.8$    &     $680$   &   $1400$  &   $3300$  \\ \hline
    
\end{tabular}
\end{center}
\end{table}
 
\begin{figure} 
\begin{center}
\includegraphics[width=.5\textwidth]{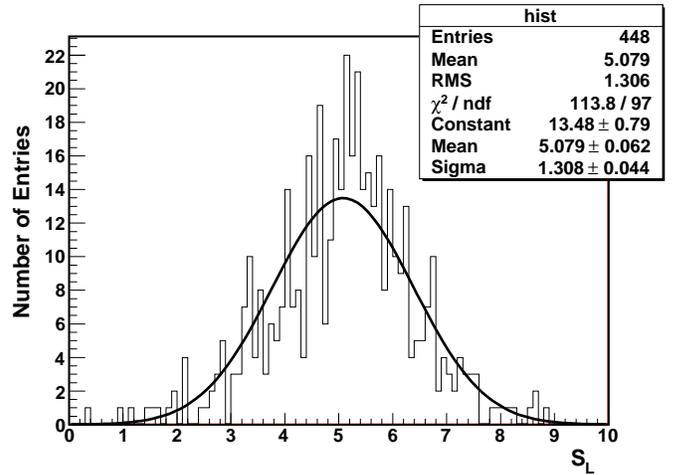}
\caption{Significance distribution $S_L$ for $M_{TS} = 1$ TeV and $\eta_f=0.2$}
\label{fig:expsig}
\end{center} 
\end{figure}

To obtain the statistical significance (Eq.4), Equation 5 is fitted to the invariant mass of the events that pass the cut described above. The fit is performed in two steps: first, using Equation 5, we fit the invariant mass for many different MC samples, and take the mean values of $M_{TS}$ and $\Gamma_{TS}$ from the fitted parameters distribution. Then, fixing $M_{TS}$ and $\Gamma_{TS}$ at their mean values, we fit the same samples again, under the signal plus background hypothesis (note that in this case, just $f_s$ is fitted), in order to obtain $L_{s+b}$. By making $f_s$ = 0 in Equation 5, which  represents the background only hypothesis, we fit the same distributions to get $L_{b}$. In the last fit the parameters $k_1$ and $k_2$ are free. As the result of many fits a $S_L$ distribution is obtained, as illustrated for the case $M_{TS} = 1 \, \textnormal{TeV}$ and $\eta_{f} = 0.2$ in Figure ~\ref{fig:expsig}. The mean value of $S_L$ distribution is taken as the signal significance for a given luminosity. The value of  $S_{L}$ must be greater than 5 for claiming a discovery. 

\begin{figure} 
\begin{center}
\includegraphics[width=.5\textwidth]{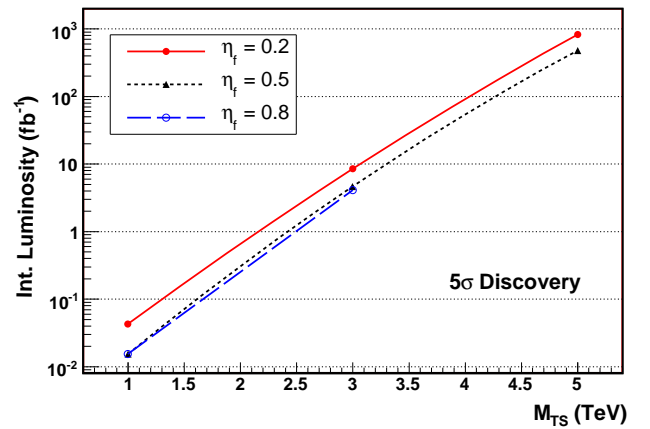}
\caption{Minimum integrated luminosity needed for a 5 $\sigma$ discovery}
\label{fig:descoberta}
\end{center} 
\end{figure}

Fig.~\ref{fig:descoberta} shows the integrated luminosity needed for a 5$\sigma$ discovery as a function of torsion mass for three different values of torsion-fermion couplings.  From this plot, we obtain the following conclusions: i) a very low integrated luminosity, of order of $100\, \textnormal{pb}^{-1}$, is needed for discovering a torsion of 1 TeV mass. It means that such signal could be observed at very early days of LHC operation, even in a regime of low luminosity; ii) for a torsion mass of 3 TeV, the discovery would be possible with an integrated luminosity around $10\,\textnormal{fb}^{-1}$, or  approximately 1 year of LHC operation in low luminosity; iii) finally, around 10 years of operation in high luminosity would be needed for the observation of a 5 TeV mass resonance. However, for the case $M_{TS} = 5$ TeV and $\eta_f = 0.8$, no  peak can be observed at LHC due to the huge torsion width ($\Gamma_{TS} = 1.9\, \textnormal{TeV})$, although some excess above the DY background can still be identified. For this reason, this particular point is not included in Fig.~\ref{fig:descoberta}.

\section{Torsion Nature Determination}

We obtain now the minimum integrated luminosity needed to discriminate between the axial and the vectorial models with 95\% C. L. as a function of the torsion mass for different $\eta_f$ values.  Assuming that a torsion signal was found at LHC and its mass and width were already experimentally determined, we have considered a mass window for the dielectron invariant mass between  $M_{TS}/2$ and $M_{TS}-\Gamma_{TS}$, where $M_{TS}$ and $\Gamma_{TS}$ are the fitted values for the torsion mass and the torsion width, respectively.  This is the region where the interference terms dominate the process.

\begin{figure} 
\begin{center}
\includegraphics[width=.5\textwidth]{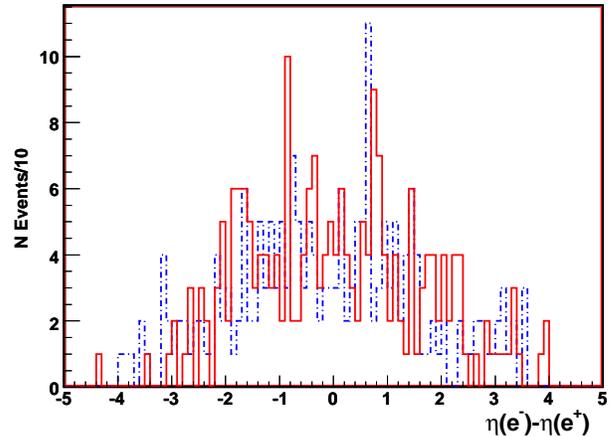}
\caption{Pseudorapidity difference between the electron and the positron for axial (solid line) and vectorial (dotted) models, for $M_{TS} =$ 3 TeV and $\eta_f =$ 0.2 for an integrated luminosity of 165 fb$^{-1}$.}
\label{fig:etadif}
\end{center} 
\end{figure}

\begin{figure} 
\begin{center}
\includegraphics[width=.5\textwidth]{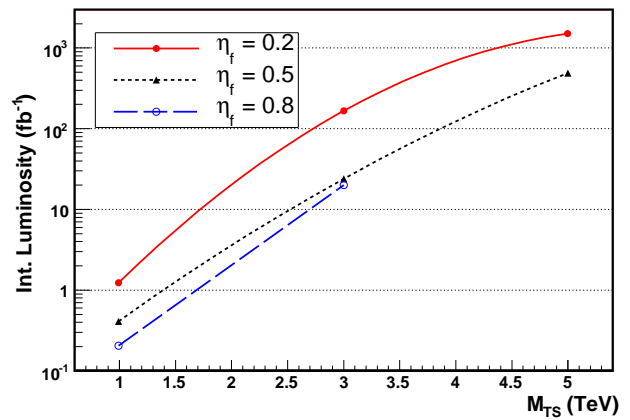}
\caption{Minimum integrated luminosity needed for 95\% C.L. discrimination between the axial and the vectorial models using the electron-positron pseudorapidity difference distribution.}
\label{fig:discriminate}
\end{center} 
\end{figure}

The pseudorapidity difference $\Delta \eta$ between the electron and the positron can be used to distinguish between the axial and the vectorial torsion models.  Due to the energy cuts on the final electron energies($E_e>5 GeV$) $\Delta \eta$ is Lorentz invariant. Fig.\ref{fig:etadif} shows the $\Delta \eta$ distribution for $M_{TS} = 3$ TeV and $\eta_f = 0.2$ for an integrated luminosity of 165 $fb^{-1}$.  In order to quantify and test the statistical compatibility of the two histograms and since either can have bins with few or even zero contents, 
we use the $\chi^2-function$ ~\cite{Marroquim} of $\Delta\eta$ distributions to calculate the minimum integrated luminosity that yields a 95\% C.L. discrimination between the axial and vectorial models.  Fig.\ref{fig:discriminate} shows our results as a function of the torsion mass for $\eta_f = $ 0.2, 0.5 and 0.8.  We can see that for $M_{TS} =$~1~TeV and $\eta_f =$ 0.8, an integrated luminosity of 0.2 fb$^{-1}$, corresponding to less than one day of LHC data-taking at high luminosity, will be enough to discriminate between axial and vector torsions, while for $M_{TS} =$ 5 TeV and $\eta_f=$ 0.2, an integrated luminosity of 1500 fb$^{-1}$, corresponding to more than 16 years of LHC data-taking at high luminosity, will be needed to achieve the torsion nature distinction.

\section{Torsion and New Neutral Gauge Boson Discrimination}

As the torsion field is characterized by a neutral gauge boson, its mass signal at LHC could be the same of a new heavy neutral resonances, like $Z^\prime$. In order to identify the signal nature, if its is found, we define an accumulated asymmetry, that is given by the forward-backward asymmetry as a function of $M_{ee}$ evaluated at increasing ranges of the final electron invariant mass $M_{ee}$. The lower range limit is at a fixed value $M_{TS}/2$ and gradually increasing the upper limit up to $M_{TS}+500$ GeV.

\beq
{\mathcal A(M_{ee})} = \frac{N_F - N_B}{N_F+N_B}\bigg|_{M_{TS}\over 2}^{M_{ee}}
\eeq

\noindent
where $N_F$ is the number of forward events, counted as the number of events with $cos\theta^{*} > 0$, and $N_B$ is the number of backward events, which have $cos\theta^{*} < 0$, where $\theta^{*}$ is the angle between the outcoming electron and the boost direction at the center of mass frame of the final electrons. 
The errors of the accumulated asymmetry become smaller as the invariant mass range increases since the total numbers of events also increases. 
 In order to compare the $Z^\prime$ and torsion fields, we have considered the $Z^\prime_{\chi}$ model, but the analysis can be equally applied to any neutral gauge boson coming from different models.

\begin{figure} 
\begin{center}
\includegraphics[width=.5\textwidth]{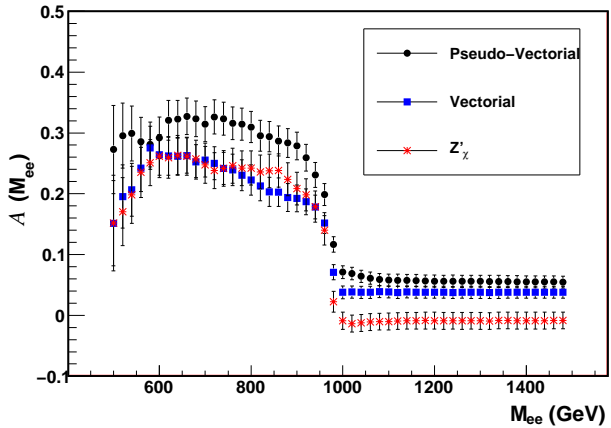}
\caption{Discrimination between the axial torsion, vectorial torsion and $Z^\prime$ using the accumulated asymmetry defined in Equation 7 for $M_{TS} = M_{Z^\prime} = 1 \, \textnormal{TeV}$ and $\eta_f = 0.2$.}
\label{fig:assimet1}
\end{center} 
\end{figure}

Figure~\ref{fig:assimet1} shows the accumulated asymmetry for the torsion fields (vectorial and pseudo-vectorial) and for $Z^\prime_{\chi}$, calculated for the case $M_{TS} = M_{Z^\prime} = 1 \, \textnormal{TeV}$, and $n_f = 0.2$ with an integrated luminosity of $20\, \textnormal{fb}^{ -1}$. We can see that the $Z^\prime$ can be easily discriminated from the pseudo-vectorial torsion in all range of mass considered. 

On the other hand, the $Z^\prime$ curve overlaps the vectorial torsion up to an invariant mass of 1 TeV, but can also be distinguished after that point. Note that it is also possible to distinguish between the vector and pseudo vector torsion fields using this variable. The discriminations are easier for greater values of $\eta_f$, since the torsion production cross section increases and therefore the errors are smaller. 

\begin{figure} 
\begin{center}
\includegraphics[width=.5\textwidth]{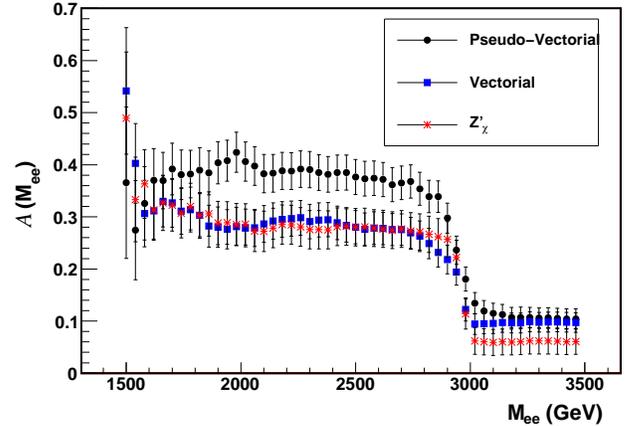}
\caption{ Same as Fig. 11 but for $M_{TS} = M_{Z^\prime} = 3 \, \textnormal{TeV}$.}
\label{fig:assimet2}
\end{center} 
\end{figure}

\begin{figure} 
\begin{center}
\includegraphics[width=.5\textwidth]{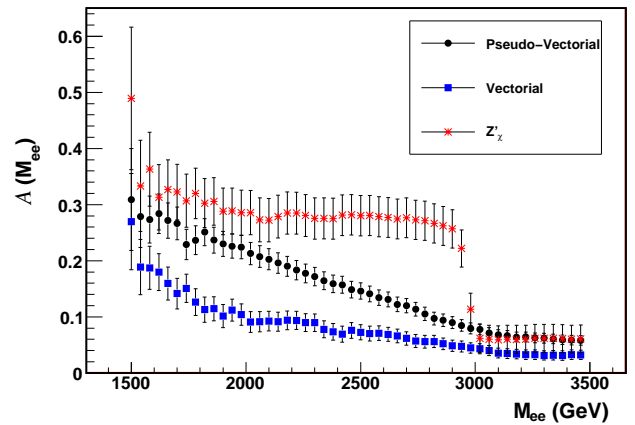}
\caption{Same as Fig. 11 but for $M_{TS} = M_{Z^\prime} = 3 \, \textnormal{TeV}$ and $\eta_f = 0.5$}
\label{fig:assimet3}
\end{center} 
\end{figure}

In Figures~\ref{fig:assimet2} and~\ref{fig:assimet3} we see the same distribution for gauge bosons mass of $3\, \textnormal{TeV}$ and for $\eta_f = 0.2$ and $0.5$, respectively. In this case, a luminosity of $900\, \textnormal{fb}^{-1}$ is needed for discrimination between the resonances. For $\eta_f = 0.2$, the accumulated asymmetry can distinguish $Z^\prime$ and pseudo-vectorial torsion, but has no power to separate the $Z^\prime$ and vectorial torsion. However, this variable shows a good performance on discrimination if $\eta_f = 0.5$, as exhibited in Figure~\ref{fig:assimet3}. 

\section{Conclusions}
  
We have studied the phenomenological consequences of a pure axial vector or pure vector torsion action at CERN LHC.  In this context, torsion is treated as a massive spin 1 particle, that will form a resonance and will decay into the SM fermions and the model parameters will be the torsion mass ($M_{TS}$) and the non minimal torsion-fermion couplings ($\eta_f$).   Using the dielectron channel we have calculated the torsion 5$\sigma$ discovery at LHC and showed that if such a resonance exists, LHC will be able to detect it after data-taking corresponding to $100\, \textnormal{pb}^{-1}$, if the torsion mass is around 1 TeV or $10\,\textnormal{fb}^{-1}$ if the torsion mass is around 3 TeV.  Unfortunately, for a torsion mass of 5 TeV and $\eta_f = 0.8$, no peak should be observed due to torsion huge width.

The initial main focus of LHC will be on discovery, but after this initial phase, focus will turn to model discrimination.  Concerning the torsion model, the discrimination will have two main points: the determination of torsion nature (axial vector or vectorial) and the distinction of the torsion field from other new particles predicted by many SM extensions, such the $Z^\prime$, that will also appear in proton-proton collisions as resonances decaying into difermions.  We showed that the pseudorapidity difference of the final electron and positron can be used to distinguish between an axial and a vectorial torsion.  As an example, the $\Delta \eta$ will be able to discriminate with $95\%$ C.L. between axial and vector torsions for $M_{TS} = 1$~TeV and $\eta_f$~=~0.8, with an integrated luminosity of 0.2 fb$^{-1}$.

Another variable capable of doing this distinction is the accumulated asymmetry, given by the forward-backward asymmetry value evaluated at different mass ranges.  But this asymmetry will require a greater integrated luminosity to achieve the distinction between different torsion natures and is better suited in the discrimination of the torsion fields from an additional neutral gauge boson predicted by the $Z^\prime_\chi$ model.  When the experimental data be available, combining the two methods will certainly increase the discrimination power of such proposed models.




{\it Acknowledgments}: The authors are very grateful to I. L. Shapiro for discussions and suggestions.  This work was supported in part by  the following Brazilian agencies: CNPq and FAPERJ. AAN thanks to the support given by the HELEN  program.


\begin{thebibliography}{99}

\bibitem{Hehl} F.W. Hehl {\it et al.} Rev. Mod. Phyd. {\bf 48}, 393 (1976).
\bibitem{Shapiro} I.L. Shapiro, Phys. Rep. {\bf 357}, 113 (2002).
\bibitem{Hammond} R.T. Hammond, Rep. Prog. Phys. {\bf 65} 599 (2002).
\bibitem{Guilherme} G. de Berredo-Peixoto, J.A. Helayel-Neto and I.L. Shapiro, JHEP {\bf 02} (2000) 003.
\bibitem{Puetzfeld} D. Puetzfeld, New Astron.Rev. 49 (2005) 59-64.
\bibitem{Gasperini} M. Gasperini, Phys. REv. Lett. 56 (1986) 2873-2876.
\bibitem{Belyaev} A.S. Belyaev and I.L. Shapiro, Phys. Lett. B {\bf 425}, 246 (1998); Nucl. Phys. {\bf B543}, 20 (1999).
\bibitem{Mahanta} U. Mahanta and S. Raychaudhuri, hep-ph/0307350.
\bibitem{Aline} A.S. Belyaev, I.L. Shapiro and M.A.B. do Vale, Phys. Rev. {\bf D 75}, 034014 (2007).
\bibitem{Kostelecky} V.A. Kostelecky, N. Russell and J.D. Tasson, Phys. Rev. Lett. {\bf 100}, 111102 (2008).
\bibitem{Kruglov} S.I. Kruglov, [arXiv:hep-ph/0804.4011].
\bibitem{Smith} W.H. Smith, [arXiv:hep-ex/0808.3131v1].
\bibitem{Rizzo} J.L. Hewett and T.G. Rizzo, Phys. Rep. {\bf 183}, 193 (1989).
\bibitem{Comphep} E.~Boos {\it et al.}  [CompHEP Collaboration],
Nucl.\ Instrum.\ Meth.\  A {\bf 534}, 250 (2004). [arXiv:hep-ph/0403113].
\bibitem{Atlas} The ATLAS Collaboration, G. Aad {\it et al.}, The ATLAS Experiment at the CERN Large Hadron Collider, 2008 JINST 3 S08003.
\bibitem{Atlas2} ATLAS Collaboration, Expected Performance of the ATLAS Experiment, Detector, Trigger and Physics, CERN-OPEN-2008-020, Geneva, 2008, to appear.
\bibitem{cousin05} R. Cousins, J. Mumford, V.Y. Value,"Detection of $Z^\prime$ Gauge Bosons in the Dimuon Decay Mode in CMS", CMS-NOTE-2005-002, Geneva, 2005.
\bibitem{cousin06} R. Cousins, J. Mumford, V.Y. Value,"Detection of $Z^\prime$ Gauge Bosons in the Dimuon Decay Mode in CMS", CMS-NOTE-2006-062, Geneva, 2006.
\bibitem{Marroquim} F.M.L. Almeida Jr., M. Barbi and M.A.B. do Vale, Nucl. Instrum. and Meth.A  {\bf 449},383-395 (2000).

\end{thebibliography}
\end{document}